\documentclass[twocolumn,showpacs,aps,prl,superscriptaddress]{revtex4}

\usepackage{graphicx}
\usepackage{dcolumn}
\usepackage{epsfig}
\usepackage{amsmath}

\newcommand{\BaBarYear}    {04}
\newcommand{\BaBarNumber}  {039}

\newcommand{\SLACPubNumber} {10861}

 \newcommand{\BaBarType}      {PUB}  

\input pubboard/babarsym
\RequirePackage{xspace}

\newcommand{\calP}{\ensuremath{{\cal P}}}

\newcommand{\pvec}{{\bf p}}

\newcommand{\acp}{\ensuremath{\calA_{ch}}}
\newcommand{\calB}{\ensuremath{{\cal B}}}

\newcommand{\DE}{\ensuremath{\Delta E}}

\newcommand{\xf}{\ensuremath{{\cal F}}}

\newcommand{\thetaT}{\ensuremath{\theta_{\rm T}}}
\newcommand{\costhr}{\ensuremath{\cos\thetaT}}

\newcommand\etal{{\it et al.}}
\newcommand{\half}{\ensuremath{{1\over2}}}

\newcommand{\bma}[1]{\boldmath{$#1$}}

\newcommand{\bfig}{\begin{figure}[htbpc!]}
\newcommand{\efig}{\end{figure}}
\newcommand\bef{\begin{figure}}
\newcommand\edf{\end{figure}}
\newcommand\dbline{\noalign{\vskip 0.10truecm\hrule}\noalign{\vskip 2pt}\noalign{\hrule\vskip 0.10truecm}}
\providecommand{\tbline}{\noalign{\vskip 0.05truecm\hrule\vskip0.05truecm}}
\newcommand\sgline{\noalign{\vskip 0.10truecm\hrule\vskip 0.10truecm}}
\newcommand\beq{\begin{equation}}
\newcommand\eeq{\end{equation}}
\newcommand\bear{\begin{array}}
\newcommand\enar{\end{array}}
\newcommand\beqa{\begin{eqnarray}}
\newcommand\eeqa{\end{eqnarray}}
\newcommand\ben{\begin{enumerate}}
\newcommand\een{\end{enumerate}}

\newcommand{\UfourS}{\ensuremath{\Upsilon(4S)}}

\newcommand{\omtoppp}{\ensuremath{{\omega\ra\pip\pim\piz}}}

\newcommand{\Kst}{\ensuremath{K^*}}
\newcommand{\Kstp}{\ensuremath{\Kstarp}}
\newcommand{\Kstz}{\ensuremath{\Kstarz}}
   \newcommand{\KstpKppiz}{\ensuremath{\Kstarp_{K^+\pi^0}}}
   \newcommand{\KstptoKppiz}{\ensuremath{\Kstarp\ra K^+\pi^0}}
   \newcommand{\KstpKspip}{\ensuremath{\Kstarp_{\Kz\pi^+}}}
   \newcommand{\KstptoKspip}{\ensuremath{\Kstarp\ra \Kz\pi^+}}
   \newcommand{\KstzKppim}{\ensuremath{\Kstarz_{K^+\pi^-}}}
   \newcommand{\KstztoKppim}{\ensuremath{\Kstarz\ra K^+\pi^-}}

\newcommand{\kzs}{\ensuremath{\KS}}

\providecommand{\fomegaKstp}{\ensuremath{\omega\Kstp}}
\providecommand{\omegaKstp}{\ensuremath{\Bp\ra\fomegaKstp}}
\newcommand{\BomegaKstp}{\ensuremath{\calB(\omegaKstp)}}
\providecommand{\romegaKstp}{\ensuremath{3.5^{+2.5}_{-2.0}\pm0.7}}

\providecommand{\ulomegaKstp}{\ensuremath{7.4}}

\providecommand{\fomegaKstz}{\ensuremath{\omega\Kstz}}
\providecommand{\omegaKstz}{\ensuremath{\Bz\ra\fomegaKstz}}
\newcommand{\BomegaKstz}{\ensuremath{\calB(\omegaKstz)}}
\newcommand{\romegaKstz}{\ensuremath{3.4^{+1.8}_{-1.6}\pm0.4}}

\providecommand{\ulomegaKstz}{\ensuremath{6.1}}

\newcommand{\fomegaKstpKspip}{\ensuremath{\omega\Kstp_{\KS \pip}}}

\newcommand{\romegaKstpKspip}{\ensuremath{3.9^{+3.7}_{-3.0}\pm1.4}}

\newcommand{\fomegaKstpKppiz}{\ensuremath{\omega \Kstp_{\Kp\piz}}}

\newcommand{\romegaKstpKppiz}{\ensuremath{3.1^{+3.4}_{-2.4}\pm0.8}}

\newcommand{\romegaKstzKppim}{\ensuremath{3.4^{+1.7}_{-1.6}\pm0.4}}

\providecommand{\ulomegaKstzKppim}{\ensuremath{6.1}}

\newcommand{\fomegarhop}{\ensuremath{\omega\rho^+}\xspace}
\newcommand{\omegarhop}{\ensuremath{\Bp\ra\fomegarhop}\xspace}
\newcommand{\Bomegarhop}{\ensuremath{\calB(\omegarhop)}\xspace}

\newcommand{\fLromegarhop}{\ensuremath{12.7^{+4.3}_{-3.9}\pm1.2}\xspace}
\newcommand{\fLRomegarhop}{\ensuremath{(\fLromegarhop)\times 10^{-6}}\xspace}

\newcommand{\Aomegarhop}{\ensuremath{0.05\pm 0.26\pm 0.02}}
\newcommand{\fLomegarhop}{\ensuremath{0.88^{+0.12}_{-0.15}}}
\newcommand{\FLomegarhop}{\ensuremath{\fLomegarhop\pm0.03}}

\newcommand{\fomegarhoz}{\ensuremath{\omega\rho^0}}
\newcommand{\omegarhoz}{\ensuremath{\Bz\ra\fomegarhoz}}
\newcommand{\Bomegarhoz}{\ensuremath{\calB(\omegarhoz)}}
\newcommand{\romegarhoz}{\ensuremath{0.59^{+1.3}_{-1.1}\pm0.35}}

\newcommand{\ulomegarhoz}{\ensuremath{3.3}\xspace}

\renewcommand{\KstpKspip}{\ensuremath{\Kstarp_{{\KS}\pi^+}}}
\renewcommand{\KstptoKspip}{\ensuremath{\Kstarp\ra \KS\pi^+}}

\renewcommand{\romegarhoz}{\ensuremath{0.6^{+1.3}_{-1.1}\pm0.4}}
\renewcommand{\Aomegarhop}{\ensuremath{5\pm 26\pm 2}}

\providecommand{\YomegaKstpKspip}{\ensuremath{11.6}}
\providecommand{\YErromegaKstpKspip}{%
                    \ensuremath{\YomegaKstpKspip^{+8.7}_{-7.2}}}

\providecommand{\YomegaKstpKppiz}{\ensuremath{5.4}}
\providecommand{\YErromegaKstpKppiz}{%
                    \ensuremath{\YomegaKstpKppiz^{+6.0}_{-4.2}}}

\providecommand{\YomegaKstzKppim}{\ensuremath{26.1}}
\providecommand{\YErromegaKstzKppim}{%
                    \ensuremath{\YomegaKstzKppim^{+12.1}_{-10.8}}}

\providecommand{\fLYomegarhop}{\ensuremath{57.7}}
\providecommand{\fLYErromegarhop}{\ensuremath{\fLYomegarhop^{+18.5}_{-16.5}}}

\providecommand{\Yomegarhoz}{\ensuremath{4.3}}
\providecommand{\YErromegarhoz}{\ensuremath{\Yomegarhoz^{+11.0}_{-9.1}}}

\renewcommand{\romegaKstpKspip}{\ensuremath{3.9^{+3.7}_{-3.0}\pm0.9}\xspace}

\renewcommand{\romegaKstpKppiz}{\ensuremath{3.1^{+3.4}_{-2.4}\pm0.9}\xspace}

\renewcommand{\romegaKstp}{\ensuremath{3.5^{+2.5}_{-2.0}\pm0.7}\xspace}
\renewcommand{\ulomegaKstp}{\ensuremath{7.4}\xspace}

\renewcommand{\romegaKstzKppim}{\ensuremath{3.4^{+1.8}_{-1.6}\pm0.4}\xspace}
\renewcommand{\ulomegaKstzKppim}{\ensuremath{6.0}\xspace}
\renewcommand{\romegaKstz}{\romegaKstzKppim\xspace}
\renewcommand{\ulomegaKstz}{\ulomegaKstzKppim\xspace}

\renewcommand{\romegarhoz}{\ensuremath{0.6^{+1.3}_{-1.1}\pm0.4}\xspace}
\renewcommand{\ulomegarhoz}{\ensuremath{3.3}\xspace}

\renewcommand{\fLromegarhop}{\ensuremath{12.6^{+3.7}_{-3.3}\pm1.6}\xspace}

\begin{document}


\begin{flushleft}
\babar-\BaBarType-\BaBarYear/\BaBarNumber \\
SLAC-PUB-\SLACPubNumber
\end{flushleft}

\title{
 \large \bf\boldmath Measurements of $B$ meson decays to $\omega\Kst$
and $\omega\rho$ 
}

%
\author{B.~Aubert}
\author{R.~Barate}
\author{D.~Boutigny}
\author{F.~Couderc}
\author{Y.~Karyotakis}
\author{J.~P.~Lees}
\author{V.~Poireau}
\author{V.~Tisserand}
\author{A.~Zghiche}
\affiliation{Laboratoire de Physique des Particules, F-74941 Annecy-le-Vieux, France }
\author{E.~Grauges-Pous}
\affiliation{Universitad Autonoma de Barcelona, E-08193 Bellaterra, Barcelona, Spain }
% ...
\collaboration{The \babar\ Collaboration}
\noaffiliation

\date{\today}

\begin{abstract}
We describe searches for $B$ meson decays to the charmless vector-vector
final states $\omega\Kst$ and $\omega\rho$ in 89
million \BB\ pairs produced 
in \epem\ annihilation at $\sqrt{s}=10.58\ \gev$.
We measure the following branching
fractions in units of $10^{-6}$:
    $\BomegaKstz = \romegaKstz ~~(<\ulomegaKstz)$,
    $\BomegaKstp = \romegaKstp ~~(<\ulomegaKstp)$,
    $\Bomegarhoz = \romegarhoz ~~(<\ulomegarhoz)$, and
    $\Bomegarhop = \fLromegarhop$.
The first error quoted is statistical, the second systematic, and the
upper limits are defined at 90\%\ confidence level.
For \omegarhop we also measure the longitudinal spin alignment fraction
$f_L=\FLomegarhop$ and charge asymmetry $\acp=(\Aomegarhop)\%$.
\end{abstract}


\maketitle

The \babar\ \cite{BaBarVV03,BaBarRhoRhoObs,BaBarRhoRhoCPdtR12}\ and Belle
\cite{BelleRhoRho0Obs}\ experiments have reported observations of $B$
meson decays to 
most of the charge states of two-body combinations of $\rho$ and \Kstar\
mesons.  Here we present the results of searches for decays to
final states with an $\omega$ meson plus a \Kstar\ or $\rho$ meson.

The decays $B\to V_1V_2$, where $V_1$ and $V_2$ are spin-one mesons,
proceed through a combination of $S$-, $P$-, and $D$-wave
amplitudes, or in the helicity basis by amplitudes $A_0$,
$A_{\pm1}$.  The subscripts give the helicities of the vector mesons.
The longitudinal $A_0$ amplitude is a linear combination of $S$-
and $D$-wave (\CP\ even),
while the transverse $A_{\pm1}$ contain all three partial waves \cite{dighe}.

The spins of the vector mesons are analyzed by their decays into
pseudoscalar mesons.  We define the helicity frame for each vector meson
as its rest frame with polar axis along the direction of the boost from
the $B$ rest frame.  Because of the limited statistics of the present
study we integrate over the azimuthal angles, exploiting the uniform
azimuthal acceptance.  The angular distribution is
\begin{eqnarray}
    &&\frac{1}{\Gamma}\frac{d^2\Gamma}{d\cos{\theta_1}d\cos{\theta_2}}
    =     \label{eq:vvAngDist} \\
&&\frac{9}{4}\left\{\frac{1}{4}(1-f_L)\sin^2{\theta_1}\sin^2{\theta_2}
     +f_L\cos^2{\theta_1}\cos^2{\theta_2}\right\}\ , \nonumber
\end{eqnarray}
where $\theta_k$ is the decay angle (defined below) in the $V_k$
helicity frame, and $f_L= 
|A_0|^2/(|A_{+1}|^2+|A_0|^2+|A_{-1}|^2)$ is the fraction
of the longitudinal spin component.  
Measurements in $B\ra\rho\rho$ decays find values close to $f_L=1$
\cite{BaBarVV03,BaBarRhoRhoObs,BaBarRhoRhoCPdtR12,BelleRhoRho0Obs}, consistent with expectation \cite{aliSuzuki}, while smaller values found
in $B\ra\phi\Kst$ \cite{BaBarBellePhiKst}\ are theoretically puzzling.
Since the different angular distributions for the longitudinal and
transverse components lead to different acceptances, the
$f_L$-dependence needs to be considered in searches as well as in detailed
studies of $B\to V_1V_2$ decays.

In the decays of \Bp\ (or the flavor-definite $\Bz\ra\Kstz X,\
\Kstz\ra\Kp\pim$) and their charge conjugates it is also of interest to
measure the direct-\CP-violating charge asymmetry $\acp \equiv
(\Gamma^--\Gamma^+)/(\Gamma^-+\Gamma^+)$ in the rates
$\Gamma^\pm=\Gamma(B^\pm\ra f^\pm)$, for each final state $f^\pm$.  

\begin{figure}[htbp]
  \begin{center}
  \includegraphics[width=1.0\linewidth]{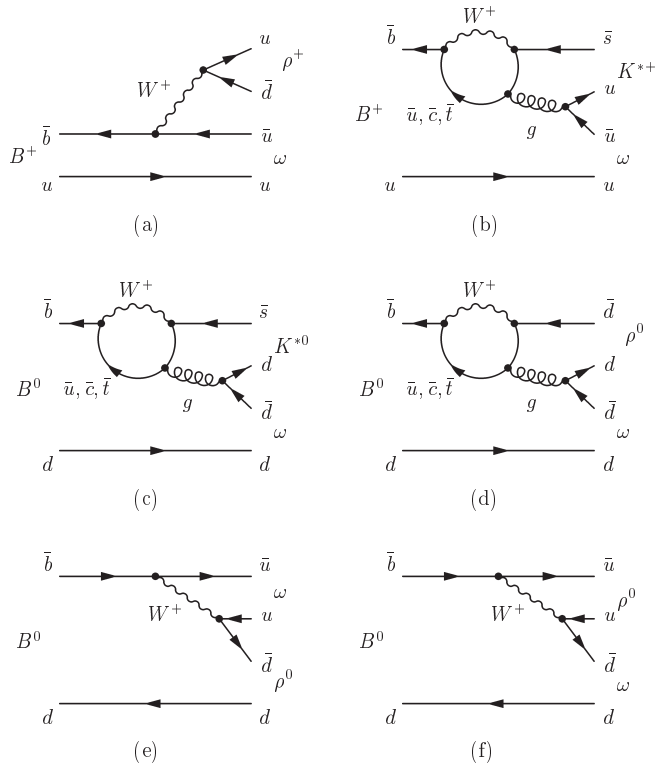}
  \end{center}
  \vspace{-3mm}
  \caption{Representative Feynman diagrams for $B\ra\omega\rho$ and
$B\ra\omega\Kstar$ decays:  (a) external tree;  (b, c) CKM-favored gluonic
penguins; (d) CKM-suppressed penguin; (e, f) destructively interfering 
color-suppressed trees.}
  \label{fig:diagram}
\end{figure}

In this study we search for the four decays
\omegaKstz, \omegaKstp, \omegarhoz, and \omegarhop \cite{CC}.  A
previously published search by CLEO 
\cite{CLEOvv}\ established 90\%\ confidence level (C.L.) upper limits on
the branching fractions of 
(23, 87, 11, and 61)$\times10^{-6}$, respectively.
Because these charmless $B$ decays involve couplings with small CKM
mixing matrix elements, several amplitudes potentially contribute with
similar strengths, as indicated in Fig.~\ref{fig:diagram}.  The
\Bp\ modes receive contributions from external tree, color-suppressed
tree, and gluonic penguin amplitudes, with the external tree (a) favored for
\omegarhop, and the penguin (b) strongly favored by CKM
couplings for \omegaKstp.  For
the \Bz\ modes there are no external tree contributions, and again,
for \omegaKstz\ the penguin (c) is CKM-favored.  For \omegarhoz\ the
color-suppressed tree amplitudes (e, f) almost cancel \cite{AliKramerLu98}\
because of the different 
isospins of the final-state mesons, leaving only a Cabibbo-suppressed 
penguin (d).  Weak exchange and annihilation amplitudes are expected to be
negligible. 

Theoretical estimates of the branching fractions for vector-vector
decays include those based on isospin relations
among various modes \cite{AtwoodSoni}, effective Hamiltonians with
factorization and specific $B$-to-light-meson form factors
\cite{Hou,AliKramerLu99,AliKramerLu98,Chen99}, and QCD factorization
\cite{ChengYang01}.  The estimated branching fractions lie in the range
$<10^{-6}$ (for \omegarhoz) to $20\times10^{-6}$ (for \omegarhop).

The results presented here are based on data collected
with the \babar\ detector~\cite{BABARNIM}
at the PEP-II asymmetric $e^+e^-$ collider~\cite{pep}
located at the Stanford Linear Accelerator Center.  An integrated
luminosity of 81.9~fb$^{-1}$, corresponding to 
$88.9\pm1.0$ million \BB\ pairs, was recorded at the $\Upsilon (4S)$
resonance (center-of-mass energy $\sqrt{s}=10.58\ \gev$).

Charged particles from the \epem\ interactions are detected, and their
momenta measured, by a combination of five layers of double-sided
silicon microstrip detectors surrounded by a 
40-layer drift chamber, both operating in the 1.5-T magnetic
field of a superconducting solenoid. We identify photons and electrons 
using a CsI(Tl) electromagnetic calorimeter (EMC).
Further charged particle identification (PID) is provided by the average energy
loss ($dE/dx$) in the tracking devices and by an internally reflecting
ring imaging Cherenkov detector (DIRC) covering the central region.

\begin{table}[btp]
\begin{center}
\caption{
Selection requirements on the invariant mass (in \mev) and decay angle
of $B$-daughter resonances. 
}
\label{tab:rescuts}
\begin{tabular}{lcc}
\dbline
State		& inv. mass	&	decay angle			\\
\sgline						
$\rho^0$	& $510 < m(\pi\pi) <1060$	&$-0.85<\cos{\theta}<0.85$\\
$\rho^+$	& $470 < m(\pi\pi) <1070$	&$-0.6<\cos{\theta}<0.85$\\
\KstzKppim,\KstpKspip &	$755 < m(K\pi)< 1035$	&$-0.85<\cos{\theta}<1.0$\\
\KstpKppiz	&$755 < m(K\pi)< 1035$	&$-0.6<\cos{\theta}<1.0$	\\
$\omega$	& $735 < m(\pi\pi\pi) <825$	&$0<|\cos{\theta}|<1.0$	\\
\piz		& $120 < m(\gamma\gamma) < 150$	&	\\
\kzs		& $488<m(\pi\pi) <508$		&	\\
\dbline
\end{tabular}
\vspace{-5mm}
\end{center}
\end{table}

We reconstruct the $B$-daughter
candidates through their decays 
$\rho^0\ra\pip\pim$, $\rho^+\ra\pip\piz$, \KstztoKppim (\KstzKppim),
\KstptoKppiz (\KstpKppiz), \KstptoKspip (\KstpKspip), \omtoppp,
$\piz\ra\gaga$, and $\kzs\ra\pip\pim$.    
Table \ref{tab:rescuts}\ lists the requirements on the invariant
mass of these particles' final states.
For the $\rho$, \Kstar, and $\omega$ invariant masses these
requirements are set loose enough to include sidebands, as these mass
values are treated as observables in the maximum-likelihood fit
described below.
For \kzs\ candidates we further require
the three-dimensional flight distance from the event primary vertex to be 
greater than three times its uncertainty.
Secondary pions and kaons in $\rho$, \Kstar, and $\omega$ candidates
are rejected if their DIRC, $dE/dx$, and EMC PID signature satisfies tight
consistency with protons or electrons, and the kaons (pions)
must (must not) have a kaon signature.

Table \ref{tab:rescuts}\ also gives the restrictions on the
\Kstar and $\rho$ helicity angles $\theta$ made to avoid regions 
of rapid acceptance variation or combinatorial background from soft
particles.  We define $\theta$ as the angle relative to the helicity
axis of: the normal to the decay plane for $\omega$, the
positively-charged (or only charged) daughter momentum for $\rho$, and
the daughter kaon momentum for \Kstar.

A $B$-meson candidate is characterized kinematically by the
energy-substituted mass $\mes=\lbrack{(\half
s+\pvec_0\cdot\pvec_B)^2/E_0^2-\pvec_B^2}\rbrack^\half$ and energy
difference $\DE = E_B^*-\half\sqrt{s}$, where the subscripts $0$ and $B$
refer to the initial \UfourS\ and to the $B$ candidate, respectively,
and the asterisk denotes the \UfourS\ frame. The resolution on \DE\
(\mes) is about 30 MeV ($3.0\ \mev$). We require $|\DE|\le0.2$ GeV and
$5.20\le\mes\le5.29\ \gev$.
The average number of candidates found per selected event is in the range
1.15 to 1.2, depending on the final state.  We choose the candidate with
the smallest value of a $\chi^2$ constructed from the deviations of the
daughter resonance masses from their expected values.

\begin{table*}[btp]
\caption{
Signal yield $Y$ and bias $Y_0$ with their statistical uncertainties,
detection efficiency $\epsilon$, daughter branching fraction product,
significance $S$ (with systematic uncertainties included), 
measured branching fraction, and 90\% C.L. upper limit for each mode.
The number of produced $B$ mesons is $(88.9\pm1.0)\times10^6$.
}
\label{tab:results}
\begin{tabular}{lccccccc}
\dbline
Mode	      	& $Y$	   &$Y_0$	&$\epsilon$ &$\prod\calB_i$ & $S$ 	&  \calB	& \calB\ U.L.	\\
		& (events) & (events)	&~~(\%)~~& (\%)	&~~($\sigma$)~~	& $(10^{-6})$	& $(10^{-6})$	\\
\dbline
\bma{\fomegaKstz}&\YErromegaKstzKppim	&$~3.2\pm1.1$	&13.2	&59	&2.2		&~\bma{\romegaKstzKppim}	&\bma{\ulomegaKstz}\\
\tbline
\fomegaKstpKspip&\YErromegaKstpKspip	&$~2.9\pm1.1$	&13.3	&20	&1.3		&\romegaKstpKspip	&---	\\
\fomegaKstpKppiz&\YErromegaKstpKppiz	&$-0.1\pm0.8$	& 6.7	&30	&1.4		&\romegaKstpKppiz	&---	\\
\bma{\fomegaKstp}&                   	&		&  	&  	&1.9	 	&~\bma{\romegaKstp}	&\bma{\ulomegaKstp}	\\
\tbline
\bma{\fomegarhoz}&\YErromegarhoz	&$-0.5\pm1.0$	&10.5	&89	&0.4		&~\bma{\romegarhoz}	&\bma{\ulomegarhoz}	\\
\tbline
\bma{\fomegarhop}&\fLYErromegarhop	&$-1.3\pm2.8$	&5.4	&89	&\bma{4.7}	&\bma{\fLromegarhop}	&---	\\
\dbline
\end{tabular}
\vspace{-5mm}
\end{table*}

Backgrounds arise primarily from random combinations of particles in
continuum $\epem\ra\qqbar$ events ($q=u,d,s,c$).  We reduce these by
selecting on the angle
\thetaT\ between the thrust axis of the $B$ candidate in the \UfourS\
frame and that of the rest of the charged tracks and neutral calorimeter
clusters in the event.  The distribution of $|\costhr|$ is sharply
peaked near $1.0$ for combinations drawn from jet-like \qqbar\ pairs,
and nearly uniform for $B$-meson decays.  The requirements, which
optimize the expected signal yield relative to its background-dominated
statistical error, are $|\costhr|<0.8$ for the \Kstar\ modes and
$|\costhr|<0.65$ for the $\rho$ modes.  In the maximum-likelihood fit we
also use a 
Fisher discriminant \xf\ \cite{fisher} that combines four variables
defined in the \UfourS\ frame: the angles with respect to the beam axis
of the $B$ momentum and $B$ thrust axis, and the zeroth and second
angular moments $L_{0,2}$ of the energy flow about the $B$ thrust axis.
The moments are defined by $ L_j = \sum_i
p_i\times\left|\cos\theta_i\right|^j,$ where $\theta_i$ is the angle
with respect to the $B$ thrust axis of track or neutral cluster $i$,
$p_i$ is its momentum, and the sum excludes the $B$ candidate daughters.

From Monte Carlo (MC) simulation \cite{geant} we estimate the residual
charmless \BB\ background to be 0.1\%\ or less of the total sample in
all cases.  To allow for contributions possibly missing in the
simulation we include a component for these in the fit described below,
with a yield free to vary.

We obtain yields, $f_L$, and \acp\ from extended unbinned 
maximum-likelihood fits with input observables \DE, \mes, \xf, and for
vector meson $k$ the mass $m_k$ and helicity-frame decay angle $\theta_k$.
For each event $i$ and hypothesis $j$ (signal, continuum background, 
\BB\ background) we define the probability density function (PDF)
\begin{equation}
\calP^i_j = \calP_j(\mes^i) \calP_j(\DE^i) 
 \calP_j(\xf^i) \calP_j(m_1^i,m_2^i,\theta^i_1,\theta^i_2).
    \label{eq:evtL}
\end{equation}
We check for correlations in the background observables beyond those
contained in 
this PDF and find them to be small.  For the signal
component, we correct for the effect of neglected correlations 
(see below).  The likelihood function is
\begin{equation}
    {\cal L} = \frac{e^{-(\sum Y_j)}}{N!} \prod_{i=1}^N
\sum_j Y_j {\cal P}_j^i\ , 
    \label{eq:totalL}
\end{equation}
where $Y_j$ is the yield of events of hypothesis $j$
found by maximizing \calL, and $N$ is the number of events in the sample.  
  
The PDF factor for the resonances in the signal takes the form
$\calP_{1,{\rm sig}}(m_1^i)\calP_{2,{\rm sig}}(m_2^i){\cal
Q}(\theta^i_1,\theta^i_2)$ with ${\cal Q}$ given by Eq.\
\ref{eq:vvAngDist}\ modified to account for detector acceptance.  For 
\qqbar\ background it is given for each resonance independently by
$\calP_{\qqbar}(m_k^i,\theta_k^i) = \calP_{\rm pk}(m_k^i)\calP_{\rm
pk}(\theta_k^i) + \calP_{\rm c}(m_k^i)\calP_{\rm c}(\theta_k^i)$,
distinguishing between true resonance ($\calP_{\rm pk}$) and
combinatorial components ($\calP_{\rm c}$).  For 
the \BB\ background we take all four mass and helicity angle observables
to be independent.  The other PDF forms are: sum of two Gaussians for
${\cal P}_{\rm sig}(\mes)$, ${\cal P}_{\rm sig}(\DE)$, and the peaking
components of ${\cal P}_j(m_k)$; a conjunction of two gaussians with
different widths below and above the peak for ${ \cal P}_j(\xf)$; and
linear or quadratic dependences for \DE, $m_k$, and helicity cosines for
\qqbar\ combinatorial background.  The \qqbar\ background in
\mes\ is described by the function
$x\sqrt{1-x^2}\exp{\left[-\xi(1-x^2)\right]}$, with
$x\equiv2\mes/\sqrt{s}$ and parameter $\xi$.

For the signal and \BB\ background components we determine the PDF
parameters from simulation.  We study large control samples of $B$ decays to
charmed final states of similar topology to verify the simulated
resolutions in \DE\ and \mes, adjusting the PDFs to account for any
differences found.  For the continuum background we use
(\mes,\,\DE) sideband data to obtain initial values, before applying the
fit to data in the signal region, and ultimately leave them free to
vary in the final fit.

Free parameters of the fit include signal and background yields,
background PDF parameters, and for the mode for which we find a
significant signal, $f_L$ and the signal and background charge
asymmetries.  For the 
fits without significant signal we vary $f_L$ by hand, basing the upper
limit on the fit with $f_L=0.9$, a choice that is consistent with a
priori expectations and gives a conservative efficiency estimate.  The
free background-PDF parameters are $\xi$ for \mes, slope for \DE, area
and slope of the combinatorial component for $m_k$, and
the peak position and lower and upper width parameters for \xf.

We evaluate possible biases from our neglect of correlations among
discriminating variables in the PDFs by fitting ensembles of simulated
experiments into which we have embedded the expected number of signal
events randomly extracted from the fully simulated MC samples.  We give
in Table \ref{tab:results}\ the values found for bias for each
mode.  Events from a weighted mixture of simulated \BB\ background
decays are included, and so the bias we measure includes the effect of
crossfeed from these modes.

In Table \ref{tab:results} we show for each decay mode the measured
branching fraction together with the quantities entering into its
computation and with its uncertainty and significance.  The statistical
error on the signal yield or branching fraction, $f_L$, and \acp\ is
taken as the change in the central value when the quantity $-2\ln{\cal
L}$ increases by one unit from its minimum value. The significance is
taken as the square root of the difference between the value of
$-2\ln{\cal L}$ (with systematic uncertainties included) for zero signal
and the value at its minimum.  For all modes except \omegarhop\ we quote
a 90\%\ C.L. upper limit, taken to be the branching fraction below which
lies 90\% of the total of the likelihood integral in the positive
branching fraction region.
In calculating branching fractions we assume that the decay rates 
of the \UfourS\ to \BpBm\ and \BzBzb\ are equal.
For decays with \Kstarp, we combine results from the two \Kstar\ decay
channels by adding the values of $-2\ln{\cal L}$, taking into account
the correlated and uncorrelated systematic errors.

\begin{figure}
\begin{center}
  \includegraphics[width=1.0\linewidth]{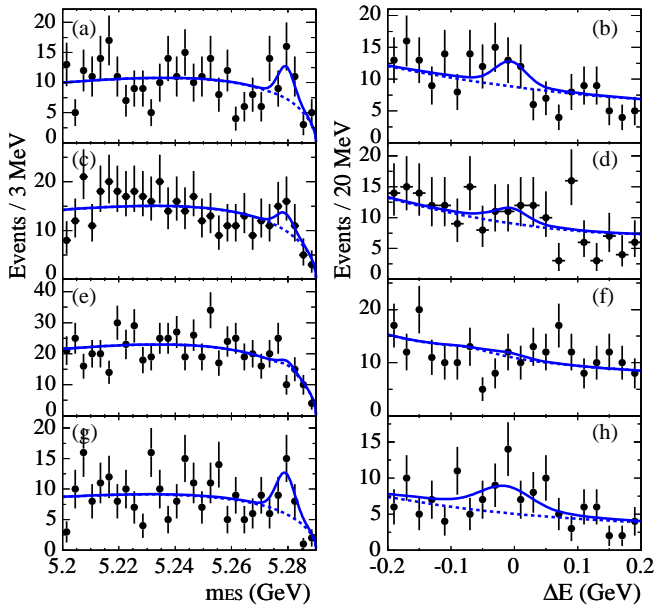}
  \caption{Projections of \mes\ (left) and \DE\
(right) with a cut on the per-event signal/total likelihood ratio for
    (a, b) \omegaKstz, (c, d) \omegaKstp, (e, f) \omegarhoz, and (g, h)
\omegarhop.  The solid (dashed) curve gives the total (background) PDF,
computed without the variable plotted, and projected 
onto the same subspace as the data.
}
  \label{fig:proj_omegaKst}
\end{center}
\end{figure}

\begin{figure}
\begin{center}
  \includegraphics[width=1.0\linewidth]{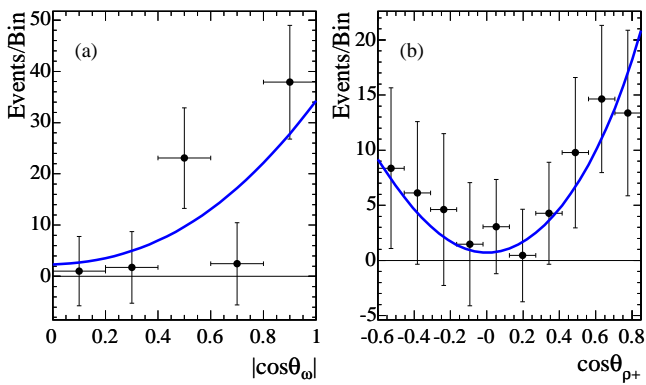}
  \vspace{-5mm}
  \caption{Background-subtracted projections of helicity cosines for
(a) $\omega$ and (b) $\rho^+$ from the fit for \omegarhop.
}
  \label{fig:sPlot_hel}
\end{center}
\end{figure}

We present in Fig.\ \ref{fig:proj_omegaKst}\ the data and PDFs projected
onto \mes\ and \DE, for subsamples enriched with a mode-dependent
threshold requirement on the ratio of signal to total likelihood
(computed without the PDF associated with the variable plotted) chosen
to optimise the significance of signal in the resulting subsample.
Fig.\ \ref{fig:sPlot_hel}\ gives background-subtracted projections
onto the helicity angle cosines for \omegarhop\ corresponding to the fit
result $f_L = \FLomegarhop$; the dominance of the term proportional to
$f_L$ in Eq.\ \ref{eq:vvAngDist}\ is evident.

\begin{figure}[htbp]
\begin{center}
 \scalebox{1.0}{
  \includegraphics[width=1.0\linewidth]{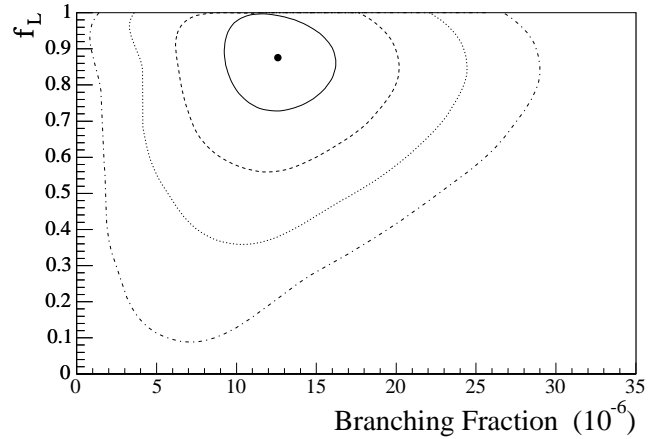}
 }
  \caption{Distribution of $\chi^2=-2\ln{\calL}(f_L,\calB)$ for \omegarhop.
The solid dot gives the central value; curves give the contours in
1-sigma steps ($\Delta\sqrt{\chi^2}=1$) out to four sigma.}  
  \label{fig:contour_omegaRho}
\end{center}
\end{figure}

The branching fraction value \calB\ given in Table \ref{tab:results}\
for \omegarhop\ comes from a direct fit with the free parameters \calB\
and $f_L$, as well as \acp.  This choice exploits the feature that
\calB\ is less correlated with $f_L$ than either the yield or
efficiency taken separately.  The behavior of $-2\ln{\calL}(f_L,\calB)$
is shown in Fig.~\ref{fig:contour_omegaRho}.

Most of the systematic uncertainties on the branching fractions arising
from lack of knowledge of the PDFs have been included in the statistical
error since most background parameters are free in the fit.  For the
signal, the uncertainties in PDF parameters are estimated from the
consistency of fits to MC and data in control modes.  Varying the
signal-PDF parameters within these errors, we estimate yield uncertainties
of 1--4 events, depending on the mode.  The uncertainty
in the fit-bias correction is taken to be half of the correction itself.
Similarly we estimate the uncertainty from modeling the \BB\ backgrounds
by taking half of the contribution of that component to the fitted
signal yield.  These additive systematic errors are dominant for the
modes with little or no signal yield.  We have also considered
backgrounds from $B$ decays to the same ultimate final state as the
signal.  The helicity angle restrctions given in Table
\ref{tab:rescuts}\ suppress $\omega\pi$ or $\omega K$ subsystems in the
region of known resonances.  For the \omegarhoz\ and \omegaKstz\ upper
limits, inclusion of the helicity-angle PDF with fixed $f_L=0.9$ reduces to a
negligible level the effect of interferences with possible $S$-wave
$\pi-(\pi,\ K)$ states.

Uncertainties in our knowledge of the efficiency, found from auxiliary 
studies, include $0.8\%\times N_t$, $2.5\%\times N_\gamma$, and 4\%\ for a
\KS\ decay, where $N_t$ and $N_\gamma$ are the number of tracks
and photons, respectively, in the $B$ candidate.  Our estimate of the
$B$-production systematic error is 1.1\%.  Published data
\cite{PDG2004}\ provide the uncertainties in the $B$-daughter product
branching fractions (1\%).  The uncertainties in the efficiency from the
event selection are 1--3\%\ for the requirement on
\costhr\ and 1\% for PID for the modes with a charged kaon.
The dependence of efficiency on $f_L$ causes uncertainties of 2--6\%\
in the $B\ra\omega K$ and \omegarhoz\ measurements.

The $0.03$ systematic error on $f_L$ for \omegarhop\ comes from
imperfect representation of 
correlations in the PDF, and is estimated from fits to fully simulated
MC samples.  From several large inclusive kaon and $B$-decay samples, we find a
systematic uncertainty for \acp\ of 2\%\ due mainly to the dependence of
reconstruction efficiency on the charge of the $\rho$-daughter charged
pion. The value of $\acp^{qq}=(-1.0\pm0.7)\%$ that we find for the
background in the \omegarhop\ fit provides confirmation of this
estimate. 

In summary, we have performed searches for the previously undetected
decays \omegaKstz, \omegaKstp, \omegarhoz\ and \omegarhop.
We observe \omegarhop\ with a significance of 4.7 $\sigma$, and
establish improved 90\% C.L. upper limits for the other modes:
\begin{eqnarray}
    \BomegaKstz &=& (\romegaKstz ~~(<\ulomegaKstz))\times10^{-6} \nonumber\\
    \BomegaKstp &=& (\romegaKstp ~~(<\ulomegaKstp))\times10^{-6} \nonumber\\
    \Bomegarhoz &=& (\romegarhoz ~~(<\ulomegarhoz))\times10^{-6} \nonumber\\
    \Bomegarhop &=& \fLRomegarhop, \nonumber
\end{eqnarray}
where the first error quoted is statistical and the second systematic.
For \omegarhop we 
also measure the longitudinal polarization fraction 
\[ f_L=\FLomegarhop \] and charge asymmetry \[\acp=(\Aomegarhop)\%.\]

We find that the longitudinal spin alignment is dominant, as for the
$\rho\rho$ modes
\cite{BaBarVV03,BaBarRhoRhoObs,BaBarRhoRhoCPdtR12,BelleRhoRho0Obs}.  The
central value of the branching fraction for \omegarhop\ is about half
of those found for $\Bp\ra\rho^+\rho^0$ and $\Bz\ra\rho^+\rho^-$.  All
of our 
branching fraction results are in general agreement within errors with
the theoretical estimates.

We are grateful for the excellent luminosity and machine conditions
provided by our \pep2\ colleagues, 
and for the substantial dedicated effort from
the computing organizations that support \babar.
The collaborating institutions wish to thank 
SLAC for its support and kind hospitality. 
This work is supported by
DOE
and NSF (USA),
NSERC (Canada),
IHEP (China),
CEA and
CNRS-IN2P3
(France),
BMBF and DFG
(Germany),
INFN (Italy),
FOM (The Netherlands),
NFR (Norway),
MIST (Russia), and
PPARC (United Kingdom). 
Individuals have received support from CONACyT (Mexico), A.~P.~Sloan Foundation, 
Research Corporation,
and Alexander von Humboldt Foundation.


\begin{thebibliography}{99}
\bibitem{CC}
Charge-conjugate reactions are included implicitly.
\end{thebibliography}
\end{document}